\documentclass[aps,prd,twocolumn,superscriptaddress,amsmath,amssymb]{revtex4}
\usepackage{xcolor,ulem}
\pdfoutput=1
\usepackage{graphicx,amsfonts,color,comment,amsmath,hyperref,float}

\begin{document}
\title{Oscillation of high-energy neutrinos from choked jets in stellar and merger ejecta}

\author{Jose Alonso Carpio}
\affiliation{Department of Physics, The Pennsylvania State University, University Park, Pennsylvania 16802, USA}
\affiliation{Center for Particle and Gravitational Astrophysics, The Pennsylvania State University, University Park, Pennsylvania 16802, USA}
\author{Kohta Murase}
\affiliation{Department of Physics, The Pennsylvania State University, University Park, Pennsylvania 16802, USA}
\affiliation{Department of Astronomy and Astrophysics, The Pennsylvania State University, University Park, Pennsylvania 16802, USA}
\affiliation{Center for Particle and Gravitational Astrophysics, The Pennsylvania State University, University Park, Pennsylvania 16802, USA}
\affiliation{Center for Gravitation and Astrophysics, Yukawa Institute for Theoretical Physics, Kyoto}

\begin{abstract}
We present a comprehensive study on oscillation of high-energy neutrinos from two different environments: blue supergiant progenitors that 
may harbor low-power gamma-ray burst (GRB) jets and neutron star merger ejecta that would be associated with short gamma-ray bursts. We incorporate the 
radiation constraint that gives a necessary condition for nonthermal neutrino production, and account for the time evolution of the jet, 
which allows us to treat neutrino oscillation in matter more accurately. For massive star progenitors, neutrino injection inside the star can 
lead to nonadiabatic oscillation patterns in the early stages  between 1~TeV and 10~TeV and is also visible in the flavor ratio.
The matter effects predict a $\nu_e$ excess in the 10~TeV -- 100~TeV range.
For neutron star merger ejecta, we 
find a similar behavior in the 100~GeV -- 10~TeV region and the oscillation may result in a $\nu_e$ excess around
1~TeV. These features, which enable us to probe the progenitors of long and short GRBs, could be seen by future neutrino 
detectors with precise flavor ratio measurements.  
We also discuss potential contributions to the diffuse neutrino flux measured by IceCube, and find parameter sets allowing choked low-power GRB jets to account for the neutrino flux in the 10~TeV--100~TeV range without violating the existing constraints.
\end{abstract}

\maketitle

\section{INTRODUCTION}
Recent observations have suggested that the population of gamma-ray bursts (GRBs) is diverse. 
Classical, high-luminosity long GRBs are typically attributed to ultrarelativistic jets from the core collapse of massive stars 
\citep[e.g.,][for reviews]{Zhang03,Modjaz11,Hjorth13,Meszaros:2015krr}. Particle acceleration in the jets will then lead to emission of gamma rays 
and perhaps production of high-energy neutrinos and ultrahigh-energy cosmic rays~\cite{Waxman95,Vietri:1995hs}.
The stacking analyses made by IceCube have shown that prompt neutrinos from GRBs do not significantly contribute to the observed diffuse neutrino flux \cite{Aartsen15a,Aartsen16}, and have given interesting constraints on the CR production in GRBs.
However, low-power GRBs (LP GRBs) such as low-luminosity GRBs (LL GRBs) with isotropic luminosities below $\sim10^{49}$ erg s$^{-1}$ 
\cite{Virgili09,Sun15} and ultralong GRBs (UL GRBs) avoid these stacking limits and may provide significant contributions to the 
diffuse flux~\cite{Murase13}. 
In particular, ``failed'' GRBs with choked jets can bypass such constraints: Unlike traditional bursts, choked GRB jets are characterized by a jet that 
does not escape the progenitor and leads to an unobservable electromagnetic signal~\cite{Meszaros01,Razzaque:2004yv,Ando:2005xi}. 
Such sources, with a population that may be much greater than classical ones, may also account for the IceCube neutrinos 
\cite{Murase13,Senno16,Senno18,Esmaili18,Tamborra16,Denton18,Denton:2018tdj}.

On the other hand, the coalescence of neutron star mergers produces gravitational waves accompanied by short GRBs (SGRBs). We can expect 
high-energy neutrino and gamma-ray emission associated with internal dissipation in relativistic outflows~\cite{Murase18,Kimura17,Biehl18}.
The SGRB jets can also be choked \cite{Nagakura14,Hamidani19} and allow for neutrino emission without accompanying photons.

As neutrinos travel to Earth, wave packet decoherence leads to an averaging out of oscillation probabilities such that the flavor ratios at injection and detection are different. In principle, measuring these ratios on the Earth can provide information on neutrino production and propagation. 
The IceCube Collaboration's first study in 2015 showed that source compositions from traditional models cannot be excluded at 68\% confidence
level \cite{Aartsen15b,Aartsen:2015ita}. Likewise, flavor ratios can be used to constrain Beyond Standard Model physics~\cite{Arguelles:2015dca,Bustamante15,Shoemaker:2015qul,Pagliaroli:2015rca,Bustamante:2016ciw,Bustamante18,Farzan19,Brdar19,Arguelles19}.

Neutrino oscillation in the context of hidden GRB jets has been studied in Refs.~\cite{Mena07,Sahu:2010ap,Razzaque10,Xiao:2015gea} both in numerical and analytical 
fashions. These previous works on the neutrino oscillation assumed the single-zone model, in which high-energy neutrinos are produced at a specific radius inside a progenitor. It was also assumed that CR acceleration occurs ad hoc, without taking into account radiation constraints that mean inefficient CR acceleration when the shock is radiation mediated~\cite{Murase13}. 
In this work, we will consider time-evolving jets, taking into account both of the radiation constraints and jet stalling conditions. This 
approach allows us to calculate time-dependent neutrino spectra as the jet propagates inside the progenitor, providing a more realistic 
calculation of high-energy neutrino production that inherently depends on the dissipation radius. We will include the radiation constraints, 
by which we can identify when the shock becomes radiation unmediated and the neutrino injection begins. On the other hand, a time-dependent 
injection site enables us to identify the density profile that neutrinos will travel through and to correctly account for the 
Mikheyev-Smirnov-Wolfenstein (MSW) effect~\cite{Wolfenstein78,Mikheyev85}, as well as the neutrino flux attenuation due to inelastic 
neutrino-nucleon scatterings. 

Here, we present a semi-analytical study of high-energy neutrino production in choked GRB jets and deal with neutrino oscillations 
numerically. For LP GRBs it is easier for the jets to become collimated inside the star, becoming slow and cylindrical 
\cite{Bromberg11,Mizuta13}. 
Under these conditions, neutrino production is more favorable in comparison to classical GRBs, where the large luminosities cause radiation-mediated shocks and inefficient CR acceleration~\cite{Murase13}. 
We also study choked SGRB jets in neutron star merger ejecta, considering internal shocks as CR acceleration sites. 

In Section II we describe the basics of relativistic jet propagation, neutrino injection and neutrino oscillations in the progenitor. 
Our results are presented in Section III, showing spectra of escaping neutrinos and observed fluxes on the Earth, as well as the
corresponding flavor ratios. We then continue to analyze in Section IV how our results can be applied to the diffuse neutrino flux seen in 
IceCube and prospects for future neutrino detectors such as IceCube-Gen2 and KM3Net.

Throughout our work we use $Q_x=Q/10^x$ and quantities are given in CGS units, unless otherwise stated.

\section{METHOD}
\subsection{Astrophysical environments}
We first describe two examples briefly. For both of our examples, we require a few common parameters: the isotropic-equivalent total luminosity $L_{\rm tot}$, the precollimated jet Lorentz factor $\Gamma_j$ and the duration $t_\text{dur}$ of the event, which are related to the jet propagation. 
In addition, the jet opening angle $\theta_j$, the magnetic energy fraction $\epsilon_B$, and the internal shock radius $r_{\rm is}$ are introduced. 
The luminosity and opening angle also determine the one-side jet luminosity, $L_j = L_{\rm tot}\theta_j^2/4$.
We consider particle acceleration associated with internal shocks. The isotropic-equivalent kinetic luminosity is given by 
$L_{\text{iso}}=\Gamma_j L_\text{tot}/\eta$, with $\eta$ being the maximum Lorentz factor and $\Gamma_j$ being the jet Lorentz factor. 
We have $L_{\text{iso}}=L_\text{tot}$ if $\Gamma_j=\eta$.

\subsubsection{Choked LP GRB jets in a massive star}
We consider a LP GRB jet as expected for UL GRBs and LL GRBs. In this environment it is possible for the jet to become collimated inside a massive star progenitor, with the collimation occurring at \cite{Bromberg11,Mizuta13}
\begin{equation}
r_\text{cs}=\left(\frac{L_j^3t^4}{c^5\theta_j^2\varrho_a^3}\right)^{1/10}\left(\frac{6\xi_h\xi_c^2}{\pi^{3/2}f_{\rm cc}\xi_a}\right)^{1/5}, 
\label{col}
\end{equation}
where $\varrho_a$ is the ambient density at $r$. The jet opening angle is assumed to be $\theta_j\sim0.1-1$. A relatively large opening angle could be realized as motivated by observations of transrelativistic supernovae~\citep[e.g.,][]{Campana:2006qe}, but instead one can consider lower-luminosity jets.
The parameters  $\xi_a=3/(3-\alpha)$ and $\xi_h=\xi_c=(5-\alpha)/3$ depend on
$\alpha=-d\ln \varrho_a/ d\ln r$, where the derivative is evaluated at the location of interest, and $f_{\rm cc}\approx0.01$ is a correction 
factor determined by numerical calculations \cite{Mizuta13}. Based on the definition of $\xi_a$, it follows that this formula is not 
applicable when the density profile falls faster than $r^{-3}$ (see Ref.~\cite{Bromberg11}). 
In this work, we use the collimation shock radius set by the cocoon pressure evaluated at the jet head radius $r_h$.
Note that the cocoon pressure is assumed to be 
constant. In more realistic situations, a pressure gradient may exist, especially $r_{\rm cs}\ll r_h$ and there are multiple collimation 
shocks that may occur at radii smaller than Eq.~(\ref{col})~\cite{Murase13}. 

Beyond the collimation shock radius $r_\text{cs}$, the jet is cylindrical and the Lorentz factor of the collimation shock is 
$\Gamma_\text{cs}\sim 1/\theta_j$. 
On the other hand, 
the jet head velocity $\beta_h$ is given by \cite{Bromberg11,Mizuta13}
\begin{equation}
\beta_h = \left(\frac{L_j}{c^5t^2\varrho_a\theta_j^4}\right)^{1/5}\left(\frac{16\xi_a}{3\pi\xi_h\xi_c^2}\right)^{1/5}.
\label{zhformula}
\end{equation}

Inside the star, shocks may be radiation mediated and photons diffuse into the upstream region. The photons are thermalized by Compton scatterings with electrons (and electron-positron pairs). Protons then become decelerated due to coupling with thermal electrons. 
If the associated Thomson optical depth is too large, the deceleration scale becomes shorter than the size of the upstream flow, leading to
inefficient CR acceleration~\cite{Murase13}.
In this work, we assume that the CR acceleration occurs at internal shocks, whose radii are limited by the collimation shock 
radius (i.e., $r_{\rm is}\leq r_{\rm cs}$).
In the case of LP GRBs, we take $r_{\rm is} = r_{\rm cs}$.
Imposing the condition $\tau_T^u\lesssim1$ to this region as the most conservative bound, we get $n'_{u}\sigma_T(r_\text{is}/\Gamma_r)\lesssim1$, where $\sigma_T$ is the Thomson cross section and $n'_u\approx L_{\text{iso}}/(4\pi r_{\rm is}^2\Gamma_j^2 m_pc^3 \Gamma_{\rm rel-is})$ is the comoving upstream electron density, assuming an $e-p$ plasma.
Here $\Gamma_r$ is the Lorentz factor of the faster shell and $\Gamma_\text{rel-is}\approx\Gamma_r/(2\Gamma_j)$ is the relative Lorentz factor between the merged shell and the fast shell (assuming fast and slow shell both have the same mass).
In terms of the LP GRB parameters, the radiation constraint~\cite{Murase13} takes the form~\footnote{There is small difference in numerical values because $\sigma_T\sim10^{-24}~{\rm cm}^2$ is used in Eqs.~(4) and (5) Ref.~\cite{Murase13}. In this work we use $\sigma_T\approx6.65\times{10}^{-25}~{\rm cm}^2$.}
\begin{eqnarray}
&&L_{\text{iso},52}r_\text{is,10}^{-1}\Gamma_{j,2}^{-3}\nonumber\\
&&\lesssim8.5\times10^{-3}~{\rm min}[\Gamma_{\rm rel-is,0.5}^2,0.32C_1^{-1}\Gamma_{\rm rel-is,0.5}^3],
\label{EfficientCRAccelerationFormula}
\end{eqnarray}
where $C\simeq10$ is a numerical factor due to the 
generation of pairs at the shock. For this work, we ignore high-energy neutrino emission produced by CRs accelerated at collimation shocks, 
as these neutrinos would be more important in the GeV-TeV region~\cite{Murase13}. 

Eq. \eqref{EfficientCRAccelerationFormula} marks the location, where efficient CR acceleration begins \cite{Murase13}. 
For successful CR injection, we need to ensure that the radiation constraint is satisfied before the jet ends at $t_\text{dur}$ (that is the GRB duration). 
In general, $t_\text{dur}$ is a free parameter; it becomes constrained by imposing the jet stalling (failed GRB) condition, namely that the breakout time $t_\text{bo}$ (when the jet head reaches the stellar radius) is longer than $t_\text{dur}$. 
For LP GRBs, this is achieved for a nonrelativistic jet head; it will also move at a near constant velocity.
We use these relations to verify that the chosen GRB parameters and density profile form bursts with the desired properties.

\begin{figure*}[t]
\centerline{
\includegraphics[width=0.5\textwidth]{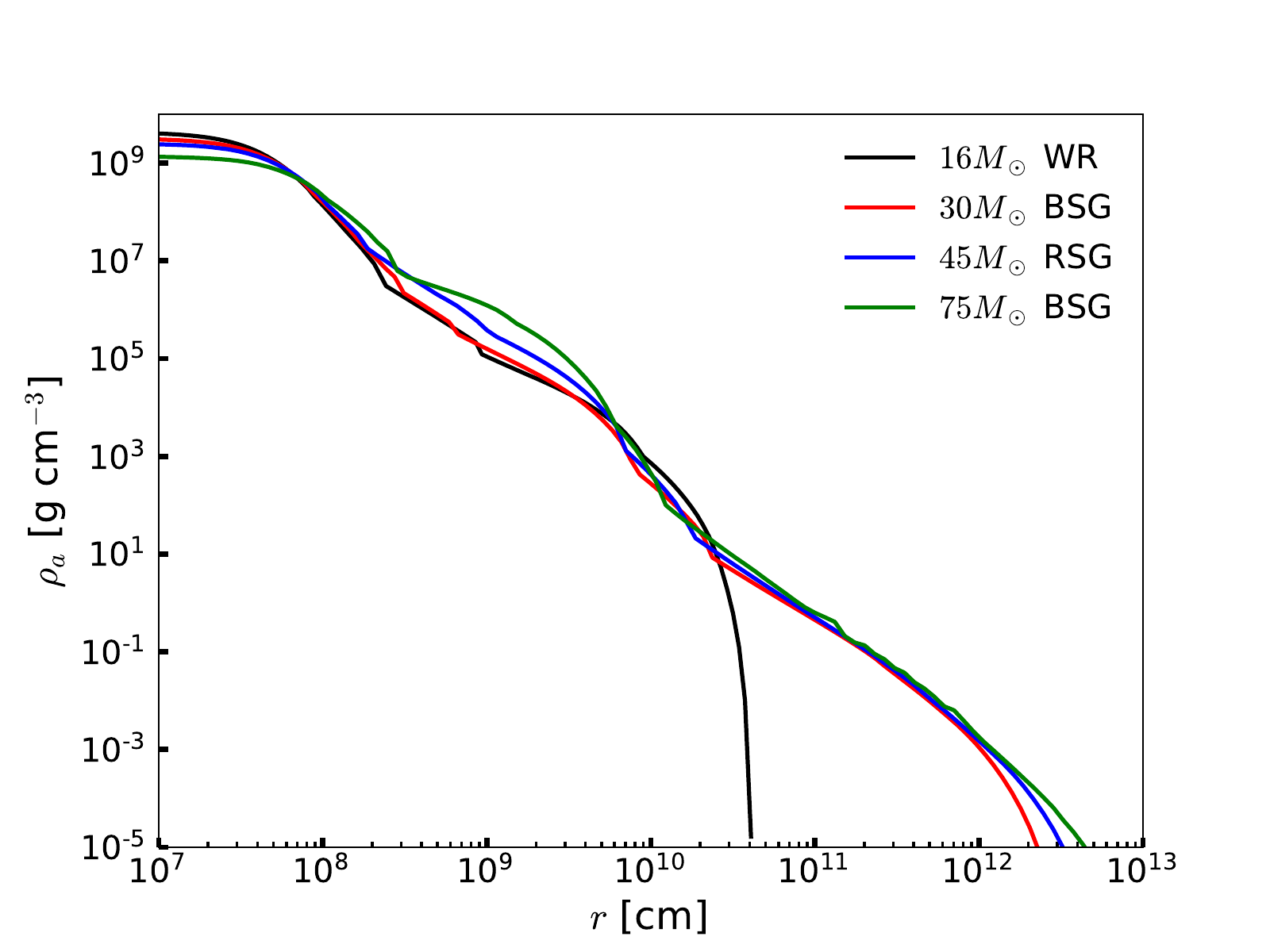}
\includegraphics[width=0.5\textwidth]{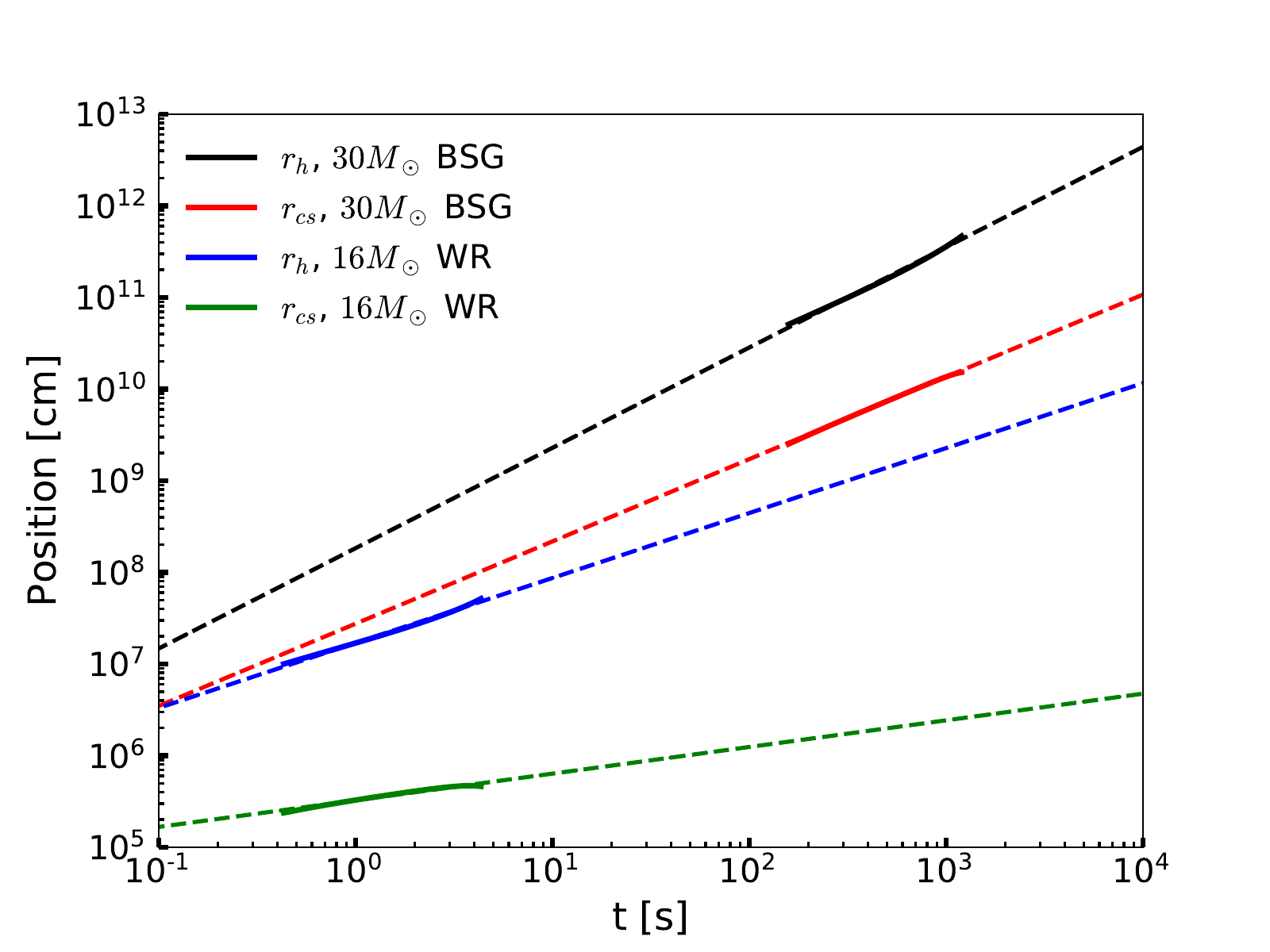}
}
\caption{
Left panel: Progenitor density profiles from \cite{Woosley02}.
Right panel: Jet head location $r_h$ and collimation shock radius $r_\text{cs}$ as a function of time. The solid lines correspond to the
	points obtained from Eq. \eqref{col} and Eq. \eqref{zhformula}, while the dashed lines are the associated extrapolations.
}
\label{Jetpropagation}
\end{figure*}

Results of the semianalytical jet propagation model are shown in Fig.~\ref{Jetpropagation}. We choose three density profiles
from \cite{Woosley02}: a 30 $M_\odot$  and $75M_\odot$ blue supergiant (BSG) and a 
45 $M_\odot$ red supergiant (RSG). We also include a 16 $M_\odot$ Wolf-Rayet (WR) profile from \cite{Woosley06}.
The radius $r_h$ is calculated using Eq. \eqref{zhformula} until we reach the point where the density profile falls off faster than $r^{-3}$. 
Beyond this point, we extrapolate to determine $r_h$. We then calculate $r_\text{cs}$ in a similar fashion, using Eq. \eqref{col}.

\subsubsection{Choked SGRB jets in merger ejecta}
For a neutron star merger, we follow the method outlined in Ref.~\cite{Kimura18} and consider the jet propagation in the merger ejecta with 
mass $M_\text{ej}$ and speed $\beta_\text{ej}$. For more detailed numerical studies see, e.g., Ref.~\cite{Hamidani19}. Jets can be launched 
through the Blandford-Znajek mechanism \cite{Blandford77} and can lead to neutrino emission by CRs accelerated at internal shocks.

We consider a time lag between the ejecta and jet production, which is given by $t_\text{lag}$, such that the ejecta radius is 
\begin{equation}
R_\text{ej} = c\beta_\text{ej}(t+t_\text{lag})
\end{equation}
and the density profile of the ejecta is wind-like as
\begin{equation}
\varrho_\text{ej} = \frac{M_\text{ej}}{4\pi R_\text{ej}^3}\left(\frac{r}{R_\text{ej}}\right)^{-2}.
\end{equation}
On the other hand, the jet head position is estimated to be
\begin{eqnarray}\nonumber
r_h &\simeq& 2.2\times 10^{10}\; L_{\text{iso},51}^{1/3}\theta_{j,-0.52}^{-2/3}M_{\text{ej},-2}^{-1/3}\beta_{\text{ej},-0.48}^{1/3}\\
& &\times t_{0.3}^{4/3}\chi_{\text{lag},0.18}\; \text{cm},
\label{EjectaRadius}
\end{eqnarray}
where $M_{\text{ej},-2} = M_\text{ej}/(0.01M_\odot)$ (this is the only exception to our definition of $Q_x$) is the ejecta mass
and $\chi_\text{lag} = 1+t_\text{lag}/t$. 
We will assume that production happens in the internal shocks, when a fast shell with Lorentz factor $\Gamma_r$ collides with a slower one of $\Gamma_s$ to form a merged shell of $\Gamma_j$. This collision occurs at the internal shock radius $r_\text{is} \simeq 8.4\times 10^9\; t_{\text{var},-4}\Gamma_{j,2.48}^2\Gamma_{\text{rel-is},0.6}^{-2}\;\text{cm}$, where $t_\text{var}$ is the variability time. Internal shocks can form either in the precollimated jet or the collimated jet; however, the Lorentz factor in the collimated jet is so low that the shock will be radiation mediated.
For this reason, as in the LP GRB case, we assume that internal shocks occur in the unshocked jet ($r_\text{is}\leq r_\text{cs}$) where the efficient CR acceleration reads
\begin{eqnarray}
&&L_{\text{iso},51}r_\text{is,10}^{-1}\Gamma_{j,2.48}^{-3}\nonumber\\
&&\lesssim2.3~{\rm min}[\Gamma_{\rm rel-is,0.5}^2,0.32C_1^{-1}\Gamma_{\rm rel-is,0.5}^3].
\end{eqnarray}
Finally, the jet stalling condition is imposed by $r_h<R_\text{ej}$.

\subsection{CR injection, timescales and neutrino production}
We assume an initial $dN'_p/d\varepsilon'_p \propto \varepsilon_p^{\prime -2} $ proton spectrum, where the primes indicate that the quantities are evaluated in
the comoving frame of the injection site (i.e., in the rest frame of the jet). 
The maximum proton energy is determined by the balance between the acceleration time $t^\prime_{p,\text{acc}} = \varepsilon'_p/(eBc)$ and its cooling time $t^\prime_\text{cool}$, while the minimum proton energy is $\Gamma_\text{rel-is}m_pc^2$. We can then normalize the 
injection spectrum such that its energy injection rate is equal to the isotropic-equivalent kinetic luminosity $L_{\text{iso}}$. 

The main pion production mechanism in GRBs is photomeson production, with a timescale $t_{p\gamma}$ given by the formula
\begin{equation}
t_{p\gamma}^{-1}(\varepsilon'_p)=c\int_0^\infty d\varepsilon' \int d\Omega' \frac{dn_\gamma'}{d\varepsilon'}(\varepsilon',\Omega')(1-\cos\theta')
\sigma_{p\gamma}\kappa_{p\gamma}
\end{equation}
where $\sigma_{p \gamma}$ is the photomeson production cross section, $\kappa_{p\gamma}$ is the proton's inelasticity, $\theta'$ is the angle between the momenta of the proton and photon and $dn'_\gamma/d\epsilon'$ is the target photon density per energy.

For choked LP GRB jets, the main target photons are generated by collimation shocks and follow a blackbody spectrum with a photon temperature of $kT'_\text{cj}\simeq 0.70~L_{\text{iso},49.5}r_{\text{cs},11.5}^{-1/2}(\theta_j/0.2)^{1/2}$keV. In the comoving frame, the photon density and the energy of each individual photon are boosted by a factor of $\Gamma_{\rm rel-cs}\approx\Gamma_j/(2\Gamma_{\rm cs})$.  
In addition, the corresponding target photon density in the inner jet is reduced by 
$\Gamma_{\rm rel-cs}[1-\exp(-\tau_\text{cj})]/\tau_\text{cj}$ because of the photon diffusion~\cite{Murase13}. 

Analogously, for choked SGRB jets, the photon density has a thermal component leaking from the collimated jet.  
Using the photon temperature $kT'_\text{cj}\simeq 9.7~\theta_{j,-0.52}^{1/2}M_{\text{ej},-2}^{1/4}\beta_{\text{ej},-0.48}^{-1/4}t_{\text{dur},0.3}^{-3/4}\chi_{\text{lag},0.18}^{-1/4}$ keV, we assume the leakage fraction to be $\tau_\text{cj}^{-1}\sim\Gamma_\text{cj}/(n'_\text{cj}\sigma_Tr_\text{cs})$, where $n'_\text{cj}\approx\Gamma_\text{rel,cs}L_{\text{iso}}/(4\pi\Gamma_j^2r_\text{cs}^2m_pc^3)$ is the density in the collimated jet. The corresponding target photon density in the inner jet becomes $\Gamma_\text{rel-cs}/\tau_\text{cj}$ times the photon densityin the collimated jet, while the energy of individual photons is also boosted by a factor $\Gamma_\text{rel-cs}$.
The non-thermal component is described by a broken power law $dn_\gamma/d\varepsilon_\gamma\propto\varepsilon_\gamma^{-\alpha_1}(\varepsilon_\gamma^{-\alpha_2})$ for  $\varepsilon_\gamma<\varepsilon_{\gamma,\text{pk}}(\varepsilon_\gamma>\varepsilon_{\gamma,\text{pk}})$, normalized such that its total energy is $U_{\gamma,\text{NT}}=\epsilon_e(\Gamma_\text{rel-is}-1)n'_\text{is}m_pc^2$, 
where $\epsilon_e$ is the fraction of thermal energy that is given to the non-thermal electrons and 
$n'_\text{is}\approx L_{\text{iso}}/(4\pi\Gamma_j^2r_\text{is}^2m_pc^3)$ is the downstream density of the internal shocks. We assume that 
the minimum (maximum) photon energy of the non-thermal component is 0.1 eV (1 MeV) and the spectral indices are $\alpha_1=0.2$ and 
$\alpha_2=2.0$~\cite{Kimura18}.

Pion production from inelastic $pp$ collisions may also have to be taken into account. The proton-proton interaction time scale is given by 
$t^\prime_{pp}=(\kappa_{pp}\sigma_{pp}n'_\text{j}c)^{-1}$. We take $\kappa_{pp}\sim 0.5$ as a constant, while the inelastic $pp$ cross 
section $\sigma_{pp}$ is parametrized by the formula given in Ref.~\cite{Kelner06}.

Using the interaction timescales $t_{pp/p\gamma}$, we can define the effective optical depth as 
\begin{equation}
	f_{p\gamma}+f_{pp} = t'_{\rm cool} ({t'}_{p\gamma}^{-1}+{t'}_{pp}^{-1}),
\end{equation}
where $t'_\text{cool}$ is found from $t_\text{cool}^{\prime -1} = \sum t^{\prime -1}$, which is a summation over all the cooling processes
in the environment of interest. For the purposes of calculating the effective optical depth, the relevant cooling processes
are $p\gamma$ and $pp$ interactions, adiabatic losses with timescale $t'_\text{ad}\approx t'_\text{dyn}\approx r_{\rm is}/c\Gamma_j$ and
synchrotron losses with timescale
\begin{equation}
t'_\text{syn} = \frac{6\pi m^4c^3}{\sigma_Tm_e^2Z^4EB'^2},
\end{equation}

for a particle of mass $m$ and energy $E$. The magnetic field in the comoving frame $B'$ satisfies the relation
\begin{equation}
\epsilon_B = \left(\frac{B'^2}{8\pi}\right)\left(\frac{L_{\text{iso}}}{4\pi r_\text{is}^2\Gamma_j^2 c}\right)^{-1},
\end{equation}
where $\epsilon_B$ is the fraction of the isotropic luminosity that is converted to magnetic field energy.

Pions and muons from $p\gamma$ interactions will lose energy as they propagate and may not be able to decay into high-energy neutrinos. 
For collimation shocks in choked long GRBs, the main pion energy loss mechanisms are synchrotron radiation and adiabatic energy loss; for 
internal shocks in choked SGRBs, we have hadronic cooling from $\pi p$ interactions in addition to the aformentioned processes. 
Muon cooling is a result of synchrotron and adiabatic losses in both astrophysical phenomena. 
The hadronic cooling timescale is ${t'}^{-1}_{\pi p}=\kappa_{\pi p}\sigma_{\pi p}n'_jc$ where we take the values 
$\kappa_{\pi p}\sim 0.8$ and $\sigma_{\pi p} \sim 5\times 10^{-26}$ cm$^{2}$ as constants for our energy range of interest. 

The pion cooling timescale is compared to its decay timescale
$t'_\text{dec}=\gamma \tau_\text{dec}$, where $\gamma$ is the Lorentz factor of the particle in the comoving frame, leading to a suppression 
factor $f_\text{sup}=1-\exp(-t'_\text{cool}/t'_\text{dec})$.
For neutrinos originating from muon decay, we require two suppression factors: one for pion cooling and another for muon cooling. The muon 
spectrum is therefore significantly suppressed with respect to the pion spectrum at high energies. We assume that the correspondence 
between the parent proton and daughter neutrino is $\varepsilon'_p\approx20 \varepsilon'_\nu (\varepsilon'_p\approx25 \varepsilon'_\nu)$ for 
$p\gamma(pp)$ interactions. In reality neutrinos from a proton with $\varepsilon'_p$ may have energies below $0.05\varepsilon'_p$ 
(or $0.04\varepsilon'_p$) due to meson and muon cooling.

\begin{figure*}[t]
\centerline{
\includegraphics[width=0.5\textwidth]{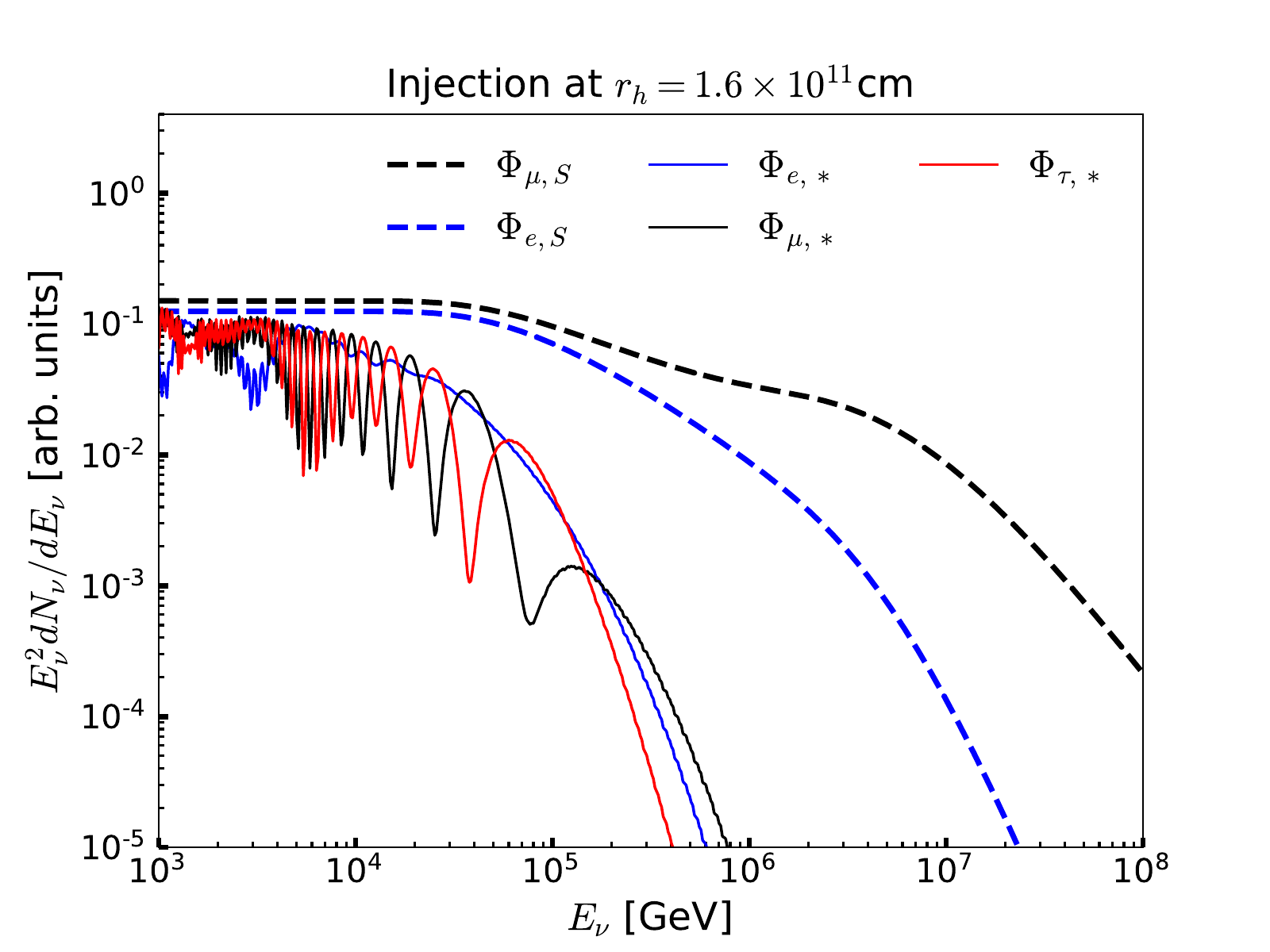}
\includegraphics[width=0.5\textwidth]{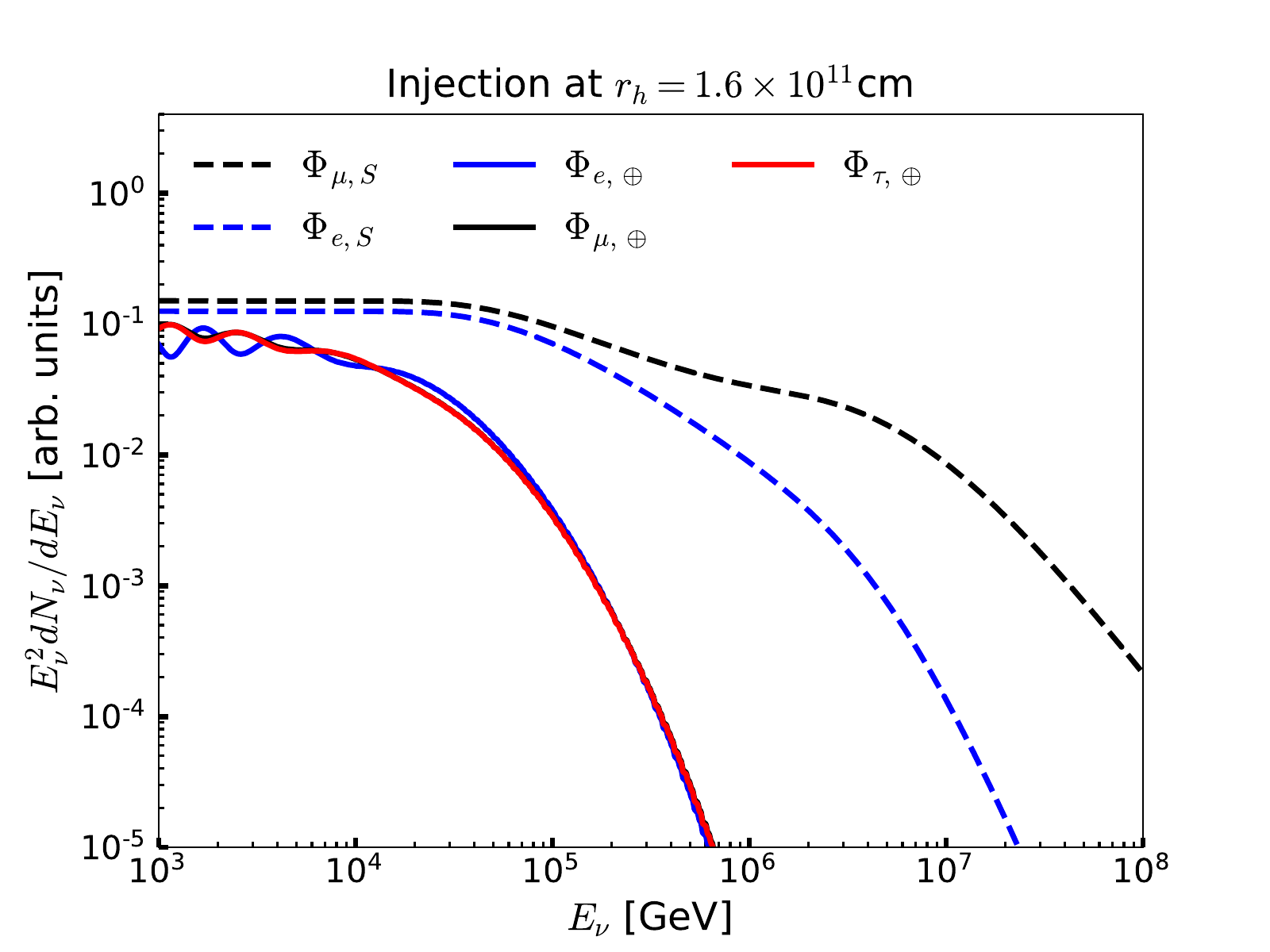}
}
\caption{Neutrino energy spectrum from a choked LP GRB jet inside a BSG. 
Left panel: Neutrino spectrum after propagating from the injection site, $r_\text{h} = 1.6\times 10^{11}$ cm, to edge of the 
source. 
Right panel: Same as left panel, showing the flux arriving at Earth after averaging out due to long distance propagation.
The proton flux is normalized such that $E_p^2 dN_p/dE_p = 1$. The $\nu_\alpha+\bar\nu_\alpha$ spectra at injection are represented by 
the dashed curves, combining contributions from $\pi$ and $\mu$ decay after accounting for cooling.}
\label{30BSGOscillation}
\end{figure*}

Meson and muon cooling modifies neutrino injection fluxes at high energies, while the production efficiency factors $f_{pp/p\gamma}$ 
modify the low-energy regions. Once we take these considerations into account, the generated neutrino spectrum ``per flavor''  in the jet 
frame is given by
\begin{equation}
{\varepsilon'_\nu}^{2}\frac{dN'_\nu}{d\varepsilon'_\nu}\approx \frac{K}{4(1+K)}{\varepsilon'_p}^2\frac{dN'_p}{d\varepsilon'_p}f_\text{sup}(f_{p\gamma}+f_{pp})
\end{equation}
where $K=1$ ($K=2$) for $p\gamma$ ($pp$) interactions, $f_\text{sup}=f_\text{sup}^\pi$ for the $\nu_\mu$ spectrum arising from pion decay and 
$f_\text{sup} = f_\text{sup}^\pi f_\text{sup}^\mu$ for the neutrinos produced as a result of muon decay. 
After we obtain the neutrino fluxes in the jet comoving frame, we perform an appropriate Lorentz boost to switch to the observer frame. 
LP GRB neutrinos are injected at $r_h$, while SGRB neutrinos are injected at $r_{\rm is}$.

\subsection{Neutrino propagation}
For neutrino propagation, we assume the following values for the oscillation parameters: $\theta_{12}=0.590,\theta_{23}=0.84,
\theta_{13}=0.15, \Delta m_{31}^2 = 2.52\times 10^{-3} \text{eV}^{2}, \Delta m_{21}^2= 7.39\times 10^{-5} \text{eV}^{2}$,
following the NuFIT 2019 oscillation fit \cite{NuFit2019}.
The effects of the CP violating phase $\delta$ are expected to be nonsignificant compared to other considerations in neutrino production,
namely the $\pi^+/\pi^-$ ratio and kaon production \cite{Blum07,Esmaili:2009dz}. 
Without these considerations, for the purpose of this work, there is little benefit in making a distinction between neutrinos and antineutrinos. We therefore treat the injection flux $\Phi_\nu+\Phi_{\bar\nu}$ as if it contained neutrinos and no antineutrinos and set $\delta=-\pi/2$
\cite{Abe18}.

For resonance effects inside the source, we use the following estimate for the $\nu_1-\nu_3$ resonance energy $E_R^H$ \cite{Razzaque10}:
\begin{equation}
E_R^H \approx \frac{\Delta m_{31}^2 \cos 2\theta_{13}}{2V} = \frac{32 \text{GeV}}{(\rho/\text{g cm}^{-3})}
\label{ResonanceEnergy}
\end{equation}
where $V = \sqrt{2}G_F n_e$ is the matter potential, $G_F$ is the Fermi constant and $n_e$ is the electron number density. The right hand
side of Eq. \eqref{ResonanceEnergy} uses the best fit values of the oscillation parameters and $n_e = Y_e \rho / m_p$, where
$\rho$ is the matter density, $m_p$ is the proton mass and $Y_e$ is the electron fraction. The electron fraction is assumed to be 1/2
both in Eq. \eqref{ResonanceEnergy} and our numerical simulations.

During propagation, neutral current (NC) interactions are considered. When dealing with charged current (CC) interactions, we are not
tracking the charged leptons formed in the process since they will have less energy and will also be quickly cooled, particularly the 
electron. 
The propagation from the injection radius to $R_*$ (or to $R_\text{ej}$ for SGRBs) is handled by nuSQuIDS \cite{Arguelles14}, giving the oscillated spectra $\Phi_{\nu_\alpha,*}$ by solving the Schr\"odinger equation for the neutrino state, within the the density matrix formalism. 
In the SGRB case, we have to keep in mind that the ejecta radius and density profile are ``time dependent'' quantities: both the location of the neutrino and time elapsed since injection have to be used to impose the neutrino escape condition.

After escaping the source, wave packet decoherence will cause subsequent vacuum oscillations to be suppressed as neutrinos make their way to Earth. 
The observed flavor flux $\Phi_{\nu_\alpha,\oplus}$ is found via
\begin{equation}
\Phi_{\nu_\alpha,\oplus} = \sum_i |U_{\alpha i}|^2 \Phi_{\nu_i,*},
\label{AveragedOscillations}
\end{equation}
where $\Phi_{\nu_i,*}$ is the neutrino flux of the vacuum mass eigenstate $i$ \cite{Razzaque10} at the edge of the progenitor. 

\section{Results on neutrino oscillation and flavor ratios at Earth}
\setlength{\tabcolsep}{0.7em}
\begin{table}[!h]\begin{center}
	Choked LP GRB jet parameters 
\begin{tabular}{cccccccc}
\hline\hline
	$L_{\rm iso,48}$ & $\theta_j$ & $\Gamma_j$ & $t_\text{dur}$ & $\epsilon_B$ & $ r_\text{is}$ & $\Gamma_\text{rel-is}$ & $\epsilon_p$\\
	$1$ & 1.0 & 50 & 1800 s & 0.1 & $r_\text{cs}$ & 4 & 0.2\\
\hline\hline
\end{tabular}
\newline\newline
Choked SGRB jet parameters
\begin{tabular}{cccccc}
\hline\hline
$L_{\rm iso,51}$ & $\theta_j$ & $\Gamma_j$ & $t_\text{dur}$ & $\epsilon_B$ & $ r_\text{is}$\\
	$1$ & 0.3 & 300 & 1.8 s & 0.1 & $8.4\times 10^9$ cm\\
\hline
$\Gamma_\text{rel-is}$ & $\epsilon_e$ & $\epsilon_p$ & $\alpha_1$ & $\alpha_2$ & $\varepsilon_{\gamma,\text{pk}}$\\
	$4$ & 0.1 & 0.2 & 0.2 & 2 & 1.7 keV\\
\hline\hline
\end{tabular}

\caption{Relevant parameters assumed for our choked LP GRB and choked SGRB models. For the special case of SGRBs, we have the additional 
	parameters $M_\text{ej}=0.02M_\odot$, $\beta_\text{ej}=0.33$ and $t_\text{lag}=1$ s.}
\label{Table1}
\end{center}\end{table}

\subsection{Applications to choked LP GRB  jets inside a blue supergiant}
The parameter set used for LP GRBs is shown in Table \ref{Table1} and the density profile corresponds to a 30 solar mass blue supergiant
(BSG) from Ref.~\cite{Woosley02}. 
By taking a variety of injection radii, we obtain the propagated spectra both at escape and on the Earth. Our choice of parameters indicate 
that efficient CR acceleration happens at $r_\text{cs}\sim 5.9\times 10^{8}$ cm at $\sim 10$~s and breakout at $\sim 4600$~s. Based on 
previous studies, which obtained the $E_\nu^2 dN_\nu / dE_\nu$ flux peak in the 100 TeV range \cite{Murase13,Xiao14}, we will study the 
spectrum in the 1 TeV - 100 PeV energy range. Throughout this energy range, pion production is highly efficient. 
Using Eq. \eqref{ResonanceEnergy}, we find that, at the injection site, $E_R^H\approx$ 6 MeV when the shock becomes radiation 
unmediated and $E_R^H\approx$ 160 TeV at $t_\text{dur}$.

We show the results of our oscillated neutrino spectra in Fig. \ref{30BSGOscillation}. The proton fluxes have been normalized so 
$E_p^2 dN_p/dE_p=1$. The observed oscillation pattern for our injection radius of $1.6\times 10^{11}$ cm is not a mere result of the MSW 
resonance: the $\nu_1-\nu_3$ resonance occurs at $< 430$ GeV at injection, below the energy range of interest. 
During propagation, we can satisfy the resonance condition in the TeV range, which may explain the peaks at $1$ TeV and 3 TeV in the 
$\nu_e$ flux.
What we mostly observe are nonadiabatic oscillations, in which oscillations are caused by the $\nu_2-\nu_3$ mixing in matter 
induced by adiabaticity breaking of the $\nu_1-\nu_3$ resonance, the so-called H-wiggles mentioned in Ref.~\cite{Razzaque10}, whose 
effect decreases as we go to energies above 10 TeV.

In the high-energy regime, we observe the attenuation of the neutrino flux as a result of both pion/muon cooling and the increase in the CC 
cross section. The effect of NC interactions slightly modifies the slope of the spectrum and we found that the changes are in the order of 
10\%. Naturally, the attenuation effects become more significant at lower injection radii; if injection occurs at $10^{10}$ cm, we would
have negligible flux at 1TeV.

On the other hand, at high energies, matter effects enhance the mass splittings inside the progenitor, effectively suppressing oscillation effects. This phenomenon typically occurs in the PeV range. If we consider the propagation close to the edge of the progenitor, where the density is the smallest, we would still find little oscillations because the vacuum oscillation lengths 
$l^\text{osc}_{jk}=4\pi E_\nu/|\Delta m^2_{jk}|\gtrsim 10^{14}$ cm are much larger than the progenitor radius. 
\begin{figure}
\includegraphics[width=0.5\textwidth]{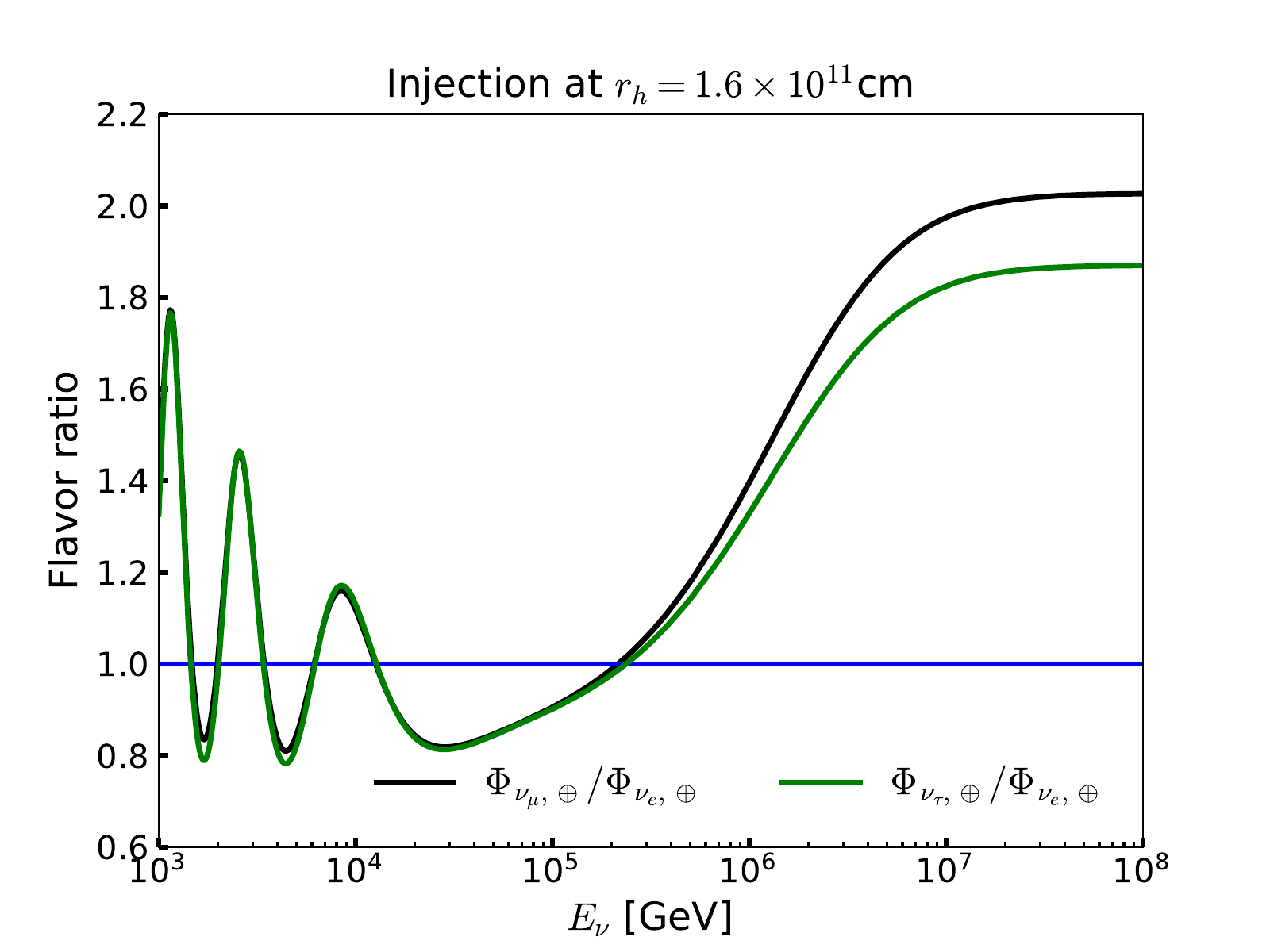}
\caption{Observed $\Phi_{\nu_\alpha}/\Phi_{\nu_e}$ flavor ratios on the Earth (i.e. oscillations are averaged out). 
Neutrino spectra are injected at 
$r_\text{h}=1.6\times 10^{11}$ cm. The blue line is a line for the (1:1) ratio and is added as a reference.}
\label{FlavorRatio1}
\vspace{-10pt}
\end{figure}

Looking at the flavor ratios, it is traditionally assumed that the neutrino spectrum at escape (for $p\gamma$ interactions) follows the 
ratio $(\nu_e:\nu_\mu:\nu_\tau)=(1:2:0)$ at escape for low energies and $(0,1,0)$ at high energies \cite{Kashti95}.
All neutrino oscillations happen in vacuum and Eq. \eqref{AveragedOscillations} takes the form
\begin{equation}
\Phi_{\alpha,\oplus} = \sum_i |U_{\alpha i}|^2 |U_{\beta i}|^2\Phi_{\beta,*},
\end{equation}
leading to the flavor ratios $(1:1.08:1.06)$ for low energies and $(1:2.03:1.87)$ at high energies. 
In our case, we inject neutrinos inside the source so matter effects will alter the low-energy ratio. We show the flavor ratios for our model in Fig. \ref{FlavorRatio1}. 
We see that nonadiabatic oscillations shown in Fig. \ref{30BSGOscillation} also induce oscillations in the flavor ratios.  

\begin{figure*}
\centerline{
\includegraphics[width=0.5\textwidth]{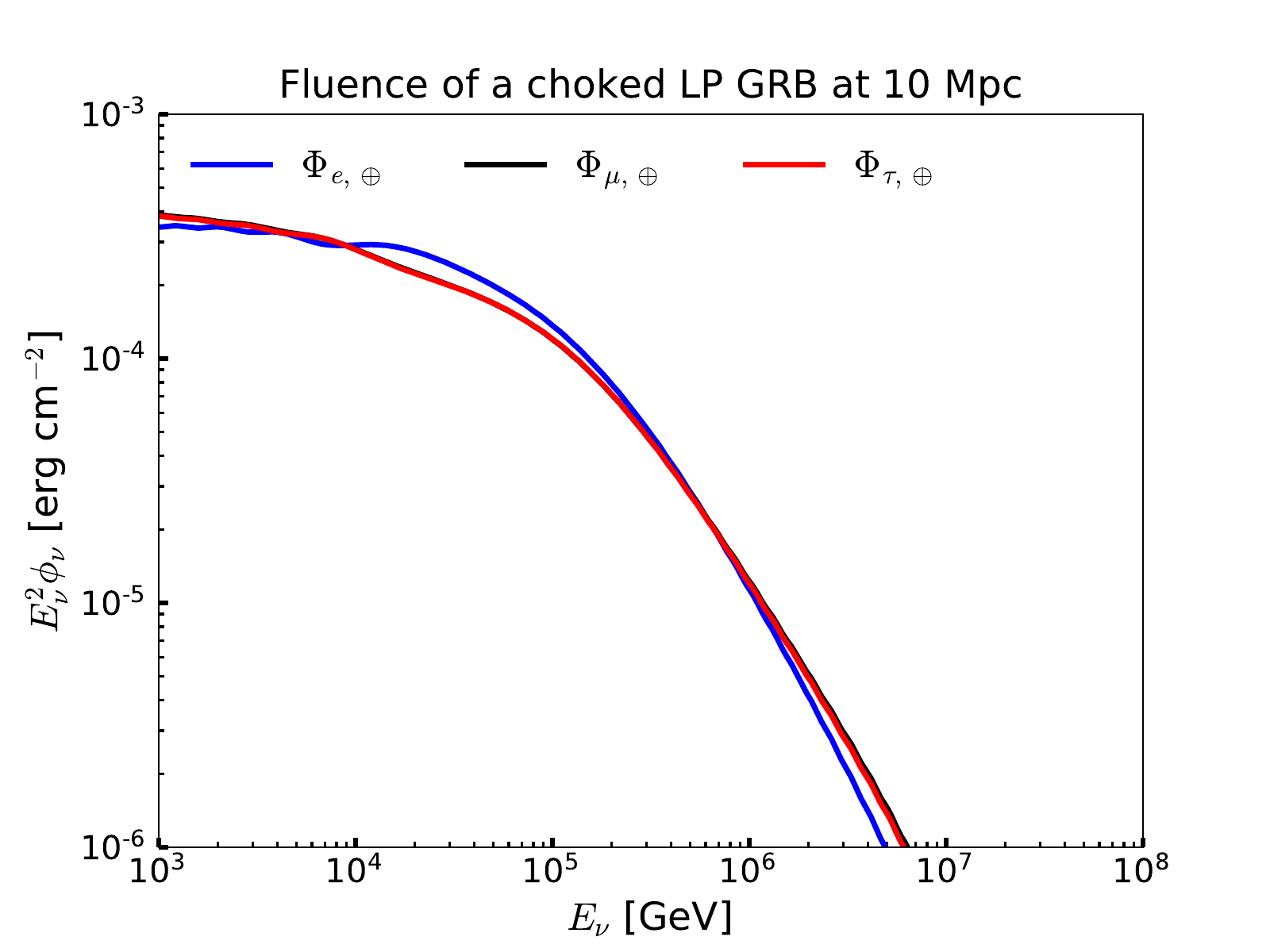}
\includegraphics[width=0.5\textwidth]{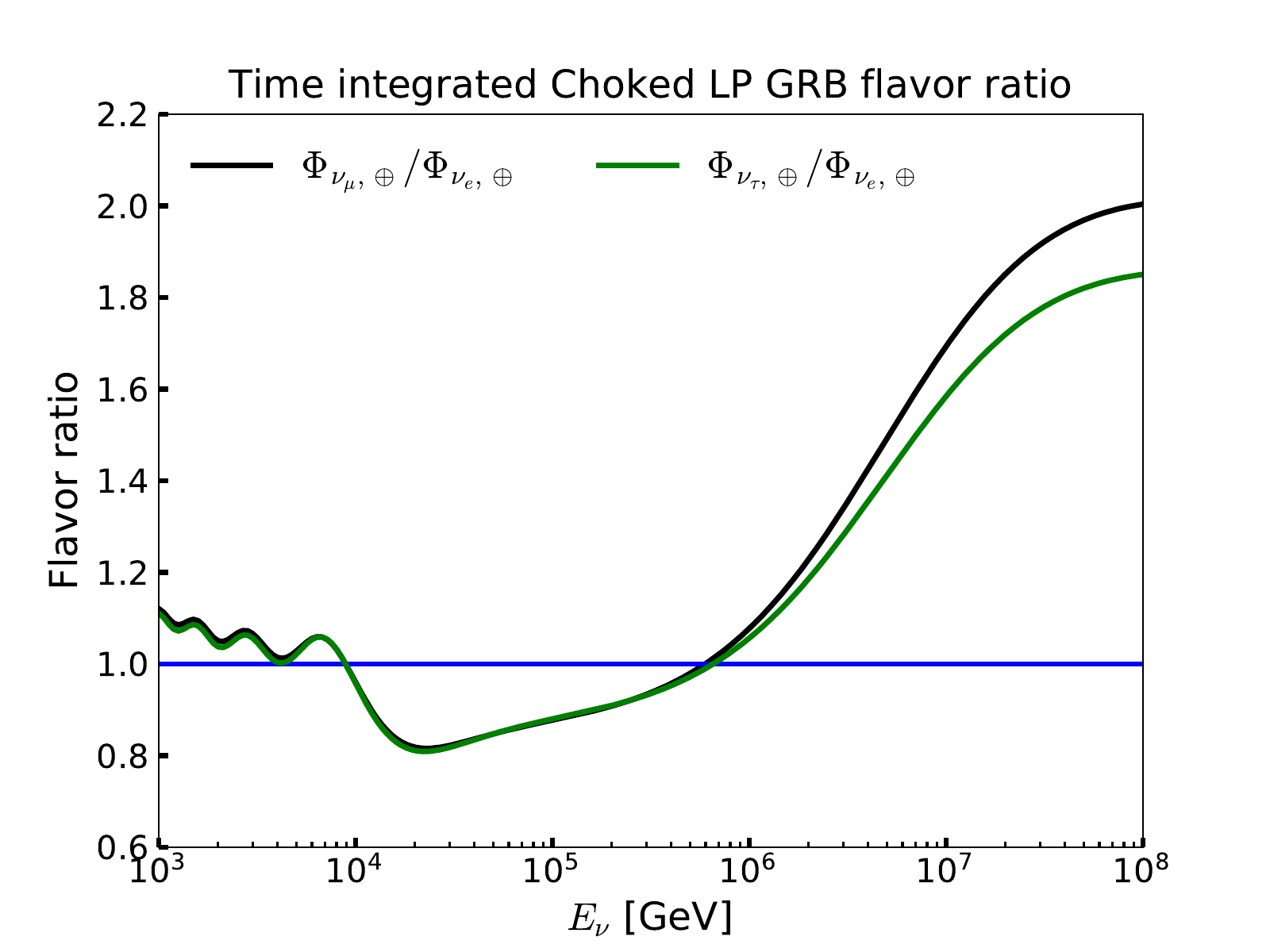}
}
\caption{Left panel: Fluence of a choked LP GRB at a distance of 10 Mpc, using the parameters of Table \ref{Table1}. Right panel: Same as left panel, but showing the flavor ratio of the fluence.}
\label{FluenceLPGRB}
\vspace{-10pt}
\end{figure*}

One feature that still persists even in the presence of matter effects is that $\Phi_{\nu_\mu}$ and $\Phi_{\nu_\tau}$ fluxes are approximately equal after averaging, for low $E_\nu$. 
The transition in the flavor ratio and the splitting between the $\nu_\mu$ and $\nu_\tau$ fluxes occurs close to 100~TeV, consistent with our theoretical expectation that the ratio approaches $(1:2.03:1.87)$ when muons are significantly cooled in the GRB. 
This transition would be hard to spot since the neutrino flux is heavily suppressed at these energies due to inelastic collisions with matter. 
Additional simulations using a 25 and 35 solar mass BSG (all other parameters fixed) show that the flavor ratio is only mildly affected by 
choosing different BSG progenitor models. Similar results hold for a red supergiant progenitor as well. We expect this because most of
the neutrino injection happens above $10^{11}$ cm, where the density profiles are similar (see Fig. \ref{Jetpropagation}).

Upon time integration up to $t_\text{dur} = 1800$~s, the flavor ratio oscillations get smeared. This can be seen in Fig. \ref{FluenceLPGRB},
where the oscillations in $\nu_e$ are less prominent. In the 1 TeV - 10 TeV range, some flavor ratio oscillations remain, with slightly more
$\nu_\mu$ and $\nu_\tau$ than $\nu_e$. In the 10 TeV - 100 TeV range we see that the $\nu_e$ excess can enhance the shower to track ratio, which could alleviate the tension between the shower and muon data (see Section~IV.B). This excess that covers a wide energy range is present because the jet is choked and matter effects are important: as we increase $t_\text{dur}$, more neutrinos are injected closer to the progenitor's edge and the fluence would approach the vacuum oscillation limit. Strong neutrino attenuation starts around 100 TeV, while at 1 PeV muon cooling occurs and the 
flavor ratio approaches $(1:2.03:1.87)$.

\subsection{Applications to choked jets LP GRB  inside a red supergiant}
In the case of a WR  star progenitor, we have $\varrho_a>10^3$ g cm$^{-3}$ until $r\sim 10^{10}$cm. 
Neutrino attenuation is important and very few neutrinos are present in the TeV range, so the only contributions come from injection close to the edge. 
We thus conclude that most of the injected neutrinos would be subject to vacuum oscillation mostly. 
If we insist on having observable matter effects, attenuation would be so strong that attempting a fit with IceCube data would inevitably overshoot the astrophysical flux in the low-energy range. Furthermore, we also get a lower bound on the allowed values of $t_\text{dur}$ if we 
are to have observable neutrinos. This restriction can be avoided if the WR star has additional surrounding material outside of its core, 
allowing for further jet propagation \cite{Nakar15}.

\begin{figure*}
\centerline{
\includegraphics[width=0.5\textwidth]{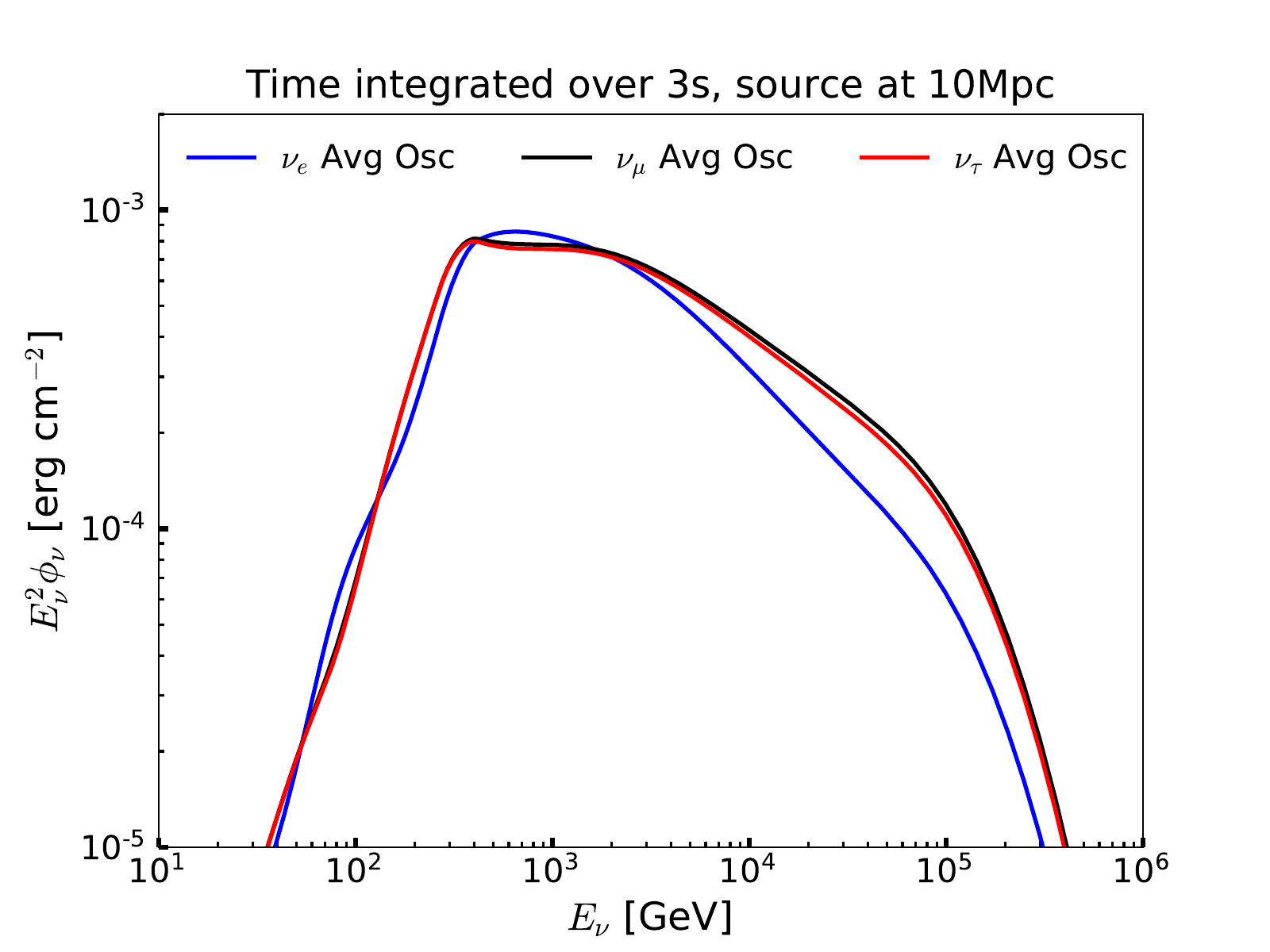}
\includegraphics[width=0.5\textwidth]{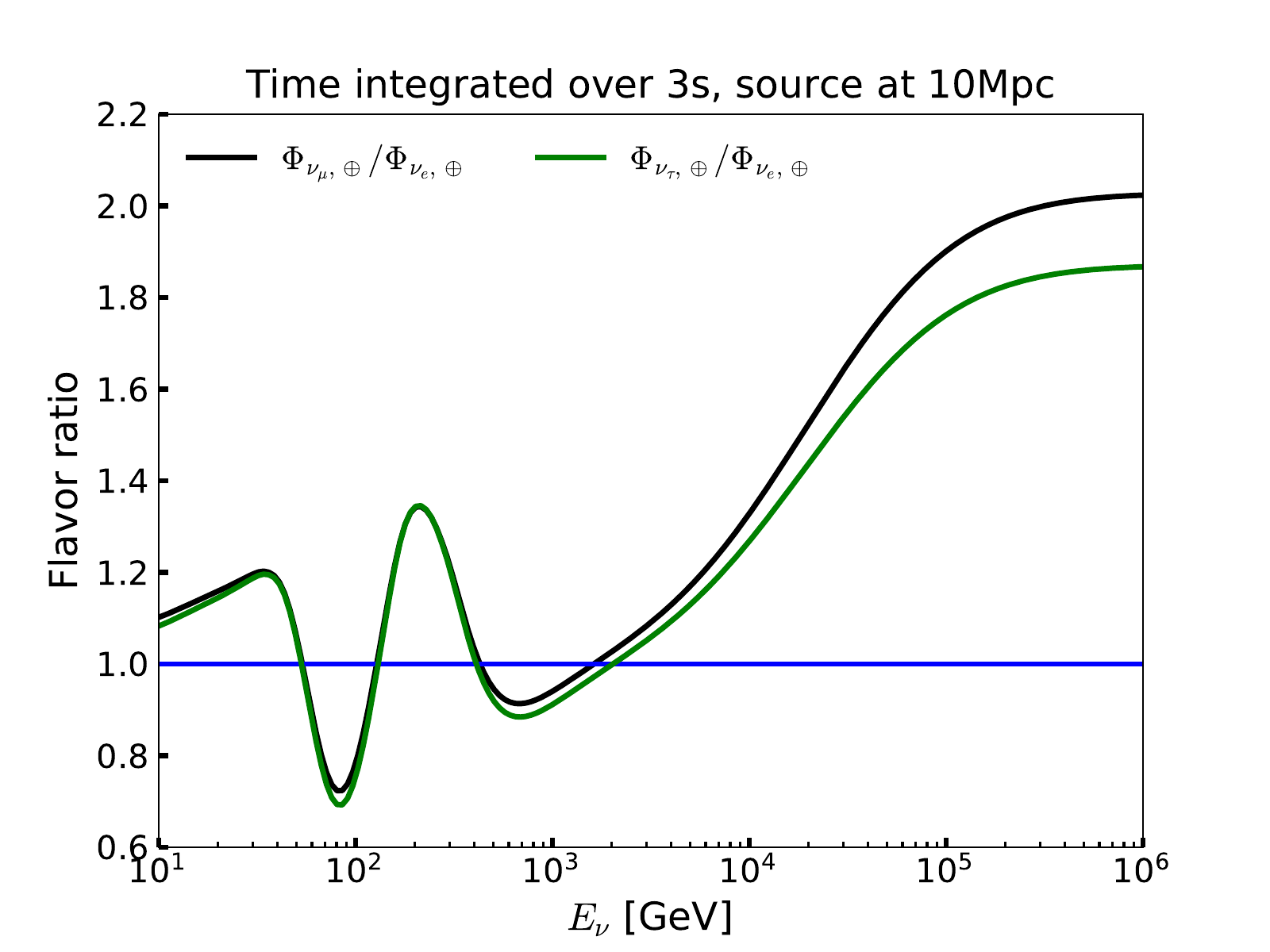}
}
\caption{Left panel: Neutrino fluence from a failed SGRB at a distance of 10~Mpc. Contributions are integrated over $t_\text{dur}=3$~s. The neutrino injection rate varies mildly over time. Right panel: Same as left panel, but showing the neutrino flavor ratio instead of the fluence.}
\label{NSMNuFlux}
\end{figure*}

\subsection{Applications to choked SGRB jets inside merger ejecta}
The parameters chosen for the choked SGRB jets are summarized in Table \ref{Table1} and the resulting oscillation pattern is shown in Fig. \ref{NSMNuFlux}. It is instructive to point out the oscillation pattern differences with respect to the LP GRB  case. 
First, we find  that the neutrino flux does not vary significantly over time; unlike LP GRBs, in which the injection begins at $\sim 10$ s, 
the constraint $r_\text{is}<r_\text{cj}$ forbids CR injection in the early phases, beginning at the neutrino onset time $t_{\rm onset}=1.7$~s 
and the duration of the neutrino injection phase is shorter in SGRBs. The mild variations in the spectra mean that the oscillations patterns 
are not smeared out after time integration. 
The $\nu_1-\nu_3$ resonance energy at the injection site occurs at 18 GeV at $t_{\rm onset}$ and $27$ GeV at $t_{\rm dur}$.

The particular parameter set that we have chosen allows for an interesting pattern to form. In the LP GRB case, the oscillation lengths are shorter than the size of the progenitor, so oscillations in the flavor ratio could be observed early, at $t=10^2$~s, but get smeared out when
integrating over $t_\text{dur}$. In the SGRB case, such flavor oscillations occur between 100~GeV and 1~TeV, which is advantageous because we can observe in Fig. \ref{NSMNuFlux} a $\sim10\%$ $\nu_e$ excess over $\nu_{\mu/\tau}$ that persists through a wide energy range after time integration. Resonance happens at $\mathcal{O}(10)$ \text{GeV}, outside our range of interest.
We also show the flavor ratio in Fig. \ref{NSMNuFlux}, showing the $\nu_e$ excess at 1TeV.
In principle, such an excess could be observed by IceCube over the 500~GeV - 30~TeV energy range.

\section{Discussion}
\subsection{Detectability of individual bursts with next-generation detectors}
It is useful to see if our predictions can be tested in future detectors such as IceCube-Gen2 and KM3Net. 
In the case of an ideal detector, for instance IceCube-Gen2, we estimate the number of events as
\begin{equation}
\mathcal{N} = \int_{E_{\nu,\text{min}}}^{E_{\nu,\text{max}}} dE_\nu {\mathcal V}(\varrho_{\rm ice}N_A)\sigma(E_\nu)\phi_\nu
\end{equation}
where $\sigma(E_\nu)$ is the neutrino-nucleon cross section, $\phi_\nu$ is the (time integrated) neutrino fluence, $\varrho_{\rm ice}$ is the ice density, $\mathcal{V}=10$~km$^3$ is the detector volume and $N_A$ is the Avogadro's constant. 
From an experimental point of view, it is often more meaningful to calculate the number of events as a function of the deposited energy. The energy deposited in the detector will depend on the neutrino flavor and on the neutrino topology. In our case, we consider fully contained events for both showers and tracks. Inclusion of partially contained events depends on selection criteria, which are not discussed in this work. 

We use the neutrino-nucleon cross sections in Ref.~\cite{Gandhi96}. The relevant shower/track channels are listed in Ref.~\cite{Blum14} and 
the deposited energy $E_\text{dep}$ for each channel is given as functions of the neutrino energy $E_\nu$ and the mean inelasticity 
$\langle y\rangle$, where the latter is obtained from Ref.~\cite{Gandhi96}. We compute the event numbers using the fluxes calculated in our 
work (referred to as ``with attenuation and oscillation''), 
as well as the fluxes obtained if we ignore matter effects and radiation constraints, while assuming that neutrino production is 
constant in time (i.e., we calculate the flux at $t_\text{dur}$ and multiply this result by $t_\text{dur}$ to find the time integrated 
fluence). We will refer to the latter scenario as the case ``without attenuation and oscillation''.

\setlength{\tabcolsep}{0.7em}
\begin{table}\begin{center}
Choked LP GRBs
\begin{tabular}{ccc}
\hline\hline
        & $E_\text{dep}>1$ TeV & $E_\text{dep}>10$ TeV\\
        \hline
Shower & 88 & 25 \\
        & (120) & (47) \\\hline
        Track & 28 & 5\\
        & (40) & (12)\\
\hline\hline
\end{tabular}
\newline\newline
Choked SGRBs

\begin{tabular}{ccc}
\hline\hline
& $E_\text{dep}>1$ TeV & $E_\text{dep}>10$ TeV\\
\hline
Shower & 65 & 10 \\
& (124) & (19) \\\hline
Track & 22 & 3\\
& (123) & (28)\\
\hline\hline
\end{tabular}
\caption{Expected number of events in IceCube-Gen2-like detectors as a result of a choked LP GRB or choked SGRB jets that occur at a distance
of 10~Mpc, assuming that the jet points to us. We use the parameters in Table \ref{Table1} and, in the case of a choked LP GRB, we use a
$30~M_\odot$ progenitor.
The event numbers are shown for two different thresholds in deposited energy. The quantities in
brackets correspond to the event numbers without attenuation and oscilation.}
\label{Table2}
\vspace{-15pt}
\end{center}\end{table}

For both our sources, we used the parameters in Table \ref{Table1}.
The results are summarized in Table \ref{Table2}, where event numbers with $E_\text{dep}>1$ TeV and $E_\text{dep}>10$ TeV are presented. 
In choked LP GRB jets, we see that the difference is less than a factor of 2 between the case with attenuation and oscillation and the 
one without. This comes from matter attenuation. The feature becomes more prominent as we increase the energy threshold for 
$E_\text{dep}$ (see blue curve in Fig. \ref{IceCubeFit}).

In the case of choked SGRB jets, we notice that a scenario without attenuation and oscillation overestimates the total number of events by a factor of $\sim 2$. By ignoring the time dependence of the problem, this case assumes neutrino emission throughout $t_\text{dur}$,
but the constraint $r_\text{is}<r_\text{cs}$ reduces this time interval by about 1/2. Without matter attenuation effects, we also
overestimate the flux and this overestimation increases with energy.
In terms of flavor ratios, we observed that the percentage of shower events increased significantly compared to the number of track events and is a feature that persists for all $E_\text{dep} >$ 1 TeV. This is caused by the $\nu_\mu\to\nu_e$ conversion above 1 TeV, reducing the number of track events, while increasing shower events. In the absence of matter effects, the $\nu_e$ flux is below $\nu_{\mu/\tau}$ flux at all energies, causing shower and track event numbers to be comparable.
Note that the non-detection of neutrinos from GRB 170817A is consistent with our model, because the SGRB jet was off-axis, preventing us from making stringent constraints from this particular event.

\subsection{Cumulative neutrino background from choked LP GRB jets}
\begin{figure}
\includegraphics[width=0.5\textwidth]{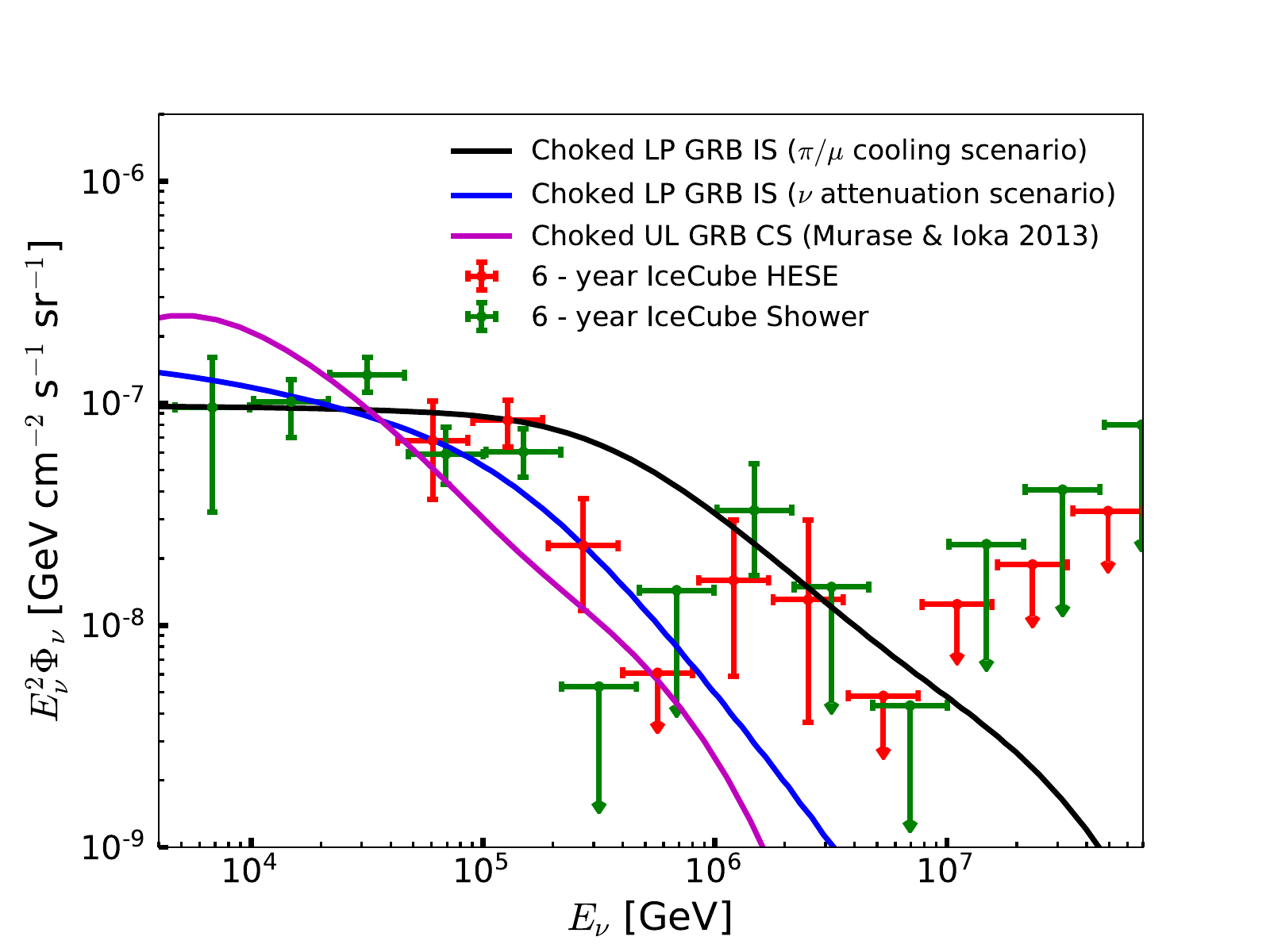}
\caption{All flavor choked LP GRB diffuse neutrino fluxes in comparison with the IceCube astrophysical neutrino spectra. 
The data from the 6-year shower analysis \cite{Aartsen20} is shown by the green bars, while the result of the 6-year HESE analysis \cite{Kopper17} is shown by the red bars. The per-flavor neutrino flux from \cite{Kopper17} was multiplied by a factor of 3 to estimate the all flavor flux. 
The $\pi/\mu$ cooling scenario uses $L_\text{iso,48}=2, \Gamma_j=70, \theta_j=0.2, t_\text{dur}=2000$ s and a 75 $M_\odot$ BSG progenitor,
while the $\nu$ attenuation scenario assumes $L_\text{iso,48}=1, \Gamma_j=50,\theta_j=1, t_\text{dur}=1800$ s and a 30 $M_\odot$ 
BSG progenitor. The remaining parameters are given in Table \ref{Table1}. For comparison, we show the spectrum of the choked UL GRB neutrinos from the collimation shock (CS) in Ref.~\cite{Murase13} but the flux is rescaled. 
}
\label{IceCubeFit}
\vspace{-10pt}
\end{figure}

We test the possibility of our oscillated neutrino spectra to match IceCube's unfolded diffuse neutrino spectrum with six years of shower data
\cite{Aartsen20} and six years of high energy starting event (HESE) data \cite{Kopper17}.  
In particular, the origin of medium-energy neutrinos has been of interest, because the multi-messenger analyses have indicated that the sources are hidden CR accelerators~\cite{Murase:2015xka,Capanema:2020rjj}, which include choked GRB jets~\cite{Murase13,He:2018lwb} and cores of active galactic nuclei~\cite{Murase:2019vdl,Kimura:2014jba}. 

We probe the $L_{\rm iso}-\Gamma_j$ space, keeping all other parameters and the progenitor model fixed. Our spectrum is time averaged, 
from the time that CR acceleration becomes efficient (see Eq. \eqref{EfficientCRAccelerationFormula}) to $t_\text{dur}$.
The normalization is left as a free parameter; we optimize it to provide a best fit to the unfolded spectrum between 10 TeV and 100 TeV.
Exploration of the parameter space is limited by the requirement $t_\text{dur}<t_\text{bo}$ and that efficient acceleration has to occur before breakout.

For this work, the normalization is set by an energy constraint that relates the total extragalactic diffuse flux to the GRB rate density as
\begin{eqnarray}\nonumber
E_\nu^2 \Phi_\nu &\sim& 4\times 10^{-8}~\text{GeV cm}^{-2}\text{s}^{-1}\text{sr}^{-1}\epsilon_p\\
	& & \times \mathcal{E}_{k,51}\left(\frac{f_{\rm cho}\rho}{1000~\text{Gpc}^{-3}\text{yr}^{-1}}\right)\left(\frac{f_z}{3}\right),\,\,\,
\end{eqnarray}
where $\mathcal{E}_k= L_{\rm iso}t_\text{dur}$ is the isotropic-equivalent kinetic energy, $f_z$ is the redshift evolution factor \cite{Waxman97,Waxman98}, $\epsilon_p$ is the energy fraction carried by CR protons, $\rho$ is the local rate density of successful LP GRBs, and $f_\text{cho}$ is the fraction of choked GRB jets compared to the successful ones. 
LP jets are preferred not only theoretically to satisfy the radiation constraints and jet stalling condition, but also observationally to be consistent with the IceCube data. 
The failed LP GRB rate density should be above $\sim60\text{ Gpc}^{-3}\text{ yr}^{-1}{(f_z/3)}^{-3}$ because a lower rate density 
contradicts the nondetection of multiplet sources \cite{Murase:2016gly,Senno:2016bso,Esmaili18,Aartsen:2018fpd}.

We find that our LP GRB jet parameters can explain the medium-energy neutrino data, which is consistent with the results of Ref.~\cite{Murase13}. Ref.~\cite{Denton:2018tdj} had difficulty in explaining the 10-100~TeV data but their parameter space is different. 
We show in Fig. \ref{IceCubeFit} the result with $L_{\rm iso,48}=1$, $\Gamma_j=50$, $t_\text{dur}\approx 1800$~s, $\theta_j=1$ and
$(\rho/1000\;\text{Gpc}^{-3}\;\text{yr}^{-1}) f_\text{cho}\sim 20$. By choosing a duration time smaller than the breakout time, we obtain a spectral cutoff due to the neutrino attenuation in the progenitor star, as expected in Ref.~\cite{Murase13}. 
For a $75M_\odot$ BSG, we choose the parameters $L_{\rm iso,48}=2, \Gamma_j=70, \theta_j=0.2$ and $t_\text{dur}\approx 2000$~s, in which the neutrino spectrum extends to the higher-energy regions. The associated rate density is $(\rho/1000\;\text{Gpc}^{-3}\;\text{yr}^{-1}) f_\text{cho}\sim 6$. In this case, neutrino attenuation is weak and the suppression is caused mainly by pion and muon cooling.
We also point out that the neutrino flavor ratio is not exactly $\approx1:1:1$ thanks to matter effects in the neutrino oscillation, and a 
$\nu_e$ excess is expected in the 10 -- 100~TeV range. This could help us explain the diffuse neutrino flux suggested by the shower analysis 
is higher than that from the upgoing muon neutrino analysis.

In both of these cases, our models are not yet constrained by the stacking limits \cite{Aartsen15a,Senno18,Esmaili18} as well as multiplet constraints \cite{Murase:2016gly,Senno:2016bso,Esmaili18,Aartsen:2018fpd}. Note that our LP GRB simulations are shown as the all-flavor diffuse neutrino fluxes; any possible flavor ratio oscillation in the low-energy region is smeared out by the summation over flavors, leaving neutrino attenuation as the relevant effect.

\section{SUMMARY AND CONCLUSIONS}
We studied neutrino production in choked  jets in LP GRBs and SGRBs.
In the case of choked LP GRB jets, we found considerable attenuation in the 10~TeV -- 100~TeV energy range by the combination of the muon
cooling and CC interactions during the initial phases of injection. In the 1~TeV -- 10~TeV region we report nonadiabatic oscillations that are not averaged out by long distance propagation; this effect is carried over to the observed flavor ratios. 
Depending on the choice of $t_\text{dur}$, a $\nu_e$ excess can be found in the neutrino fluence between 10 TeV and 100 TeV, which 
could alleviate the tension between shower and muon data.
During the later stages of injection, flavor ratio oscillations are negligible as the progenitor density decreases. The choked SGRB jet scenario allows for a 10\% $\nu_e$ excess in the TeV region, compared to the vacuum oscillation scenario where all three neutrino flavors would have an approximately equal flux, and is present over a relatively wide energy range. 

We demonstrated that $L_{\rm iso}\sim10^{48}$ erg s$^{-1}$ and $\Gamma_j\sim 50$ can provide a reasonable explanation for the IceCube 
diffuse neutrino spectrum and appropriate values for the local failed GRB rate density, with $t_\text{dur}\sim 2000$~s. For lower 
duration times, neutrino attenuation cause a flux decrease at 100 TeV without the need of cooling effects.

We discussed the detectability for future neutrino experiments such as IceCube-Gen2 and KM3Net, we found that a nearby double neutron star merger can produce 
a significant number of neutrino events at the detector. A nearby LP GRB could also yield multiple events, provided that the duration is 
sufficiently long and satisfies the choked jet constraint. In both cases, when radiation constraints and neutrino attenuation are ignored, 
neutrino events are significantly overestimated. 

The methods outlined in this manuscript can be used to provide further constraints on the parameter space, particularly on 
$\Gamma_j$ and $L_{\rm iso}$ which determine the locations where efficient acceleration begins.  On the other hand, our results can be applied to future neutrino detectors with the ability to measure the flavor ratios. Determining these ratios are important both to find the underlying neutrino production process and in finding the injection site within the progenitor, the latter being related to the transition between nonadiabatic oscillations and the suppressed oscillations inside the source. 

\begin{acknowledgements}
This work has been supported by the  Fermi GI program 111180 (K.M. and J.C.), NSF Grant No.~AST-1908689, and the Alfred P. Sloan Foundation (K.M.). 
The authors would like to thank Peter M\'esz\'aros and Irina Mocioiu for useful comments. 
 \end{acknowledgements}
 
\bibliography{JoseBibtex,kmurase}

\end{document}